\documentclass{article}

\newif\ifpreprint
\preprinttrue   

\usepackage{microtype}
\usepackage{graphicx}
\usepackage{subfigure}
\usepackage{booktabs}
\usepackage{hyperref}

\ifpreprint
  \usepackage[accepted]{mlsys2025}
\else
  \usepackage{mlsys2025}
\fi

\ifpreprint
  \makeatletter
  \renewcommand{\Notice@String}{Preprint. Not peer reviewed. Copyright \textcopyright\ 2026 held by the authors.}
  \makeatother
\fi

\mlsystitlerunning{GPUAlert: Zero-Instrumentation GPU Job Failure Diagnosis}

\usepackage{array}
\usepackage{multirow}
\usepackage{xcolor}
\usepackage{pifont}
\usepackage{tikz}
\usetikzlibrary{shapes.geometric, arrows.meta, positioning, backgrounds, fit, calc}
\usepackage{listings}

\microtypesetup{expansion=false}

\definecolor{lstbg}{gray}{0.95}
\newcommand{\cmark}{\ding{51}}
\newcommand{\xmark}{\ding{55}}

\begin{document}

\twocolumn[
\mlsystitle{GPUAlert: A Zero-Instrumentation Process-Boundary Monitor for Diagnosing GPU Training-Job Failures}

\ifpreprint
\mlsyssetsymbol{supervisor}{$\dagger$}
  \begin{mlsysauthorlist}
  \mlsysauthor{Parv Agarwal}{iitp}
  \mlsysauthor{Asif Ekbal}{iitp,supervisor}
  \end{mlsysauthorlist}
  \mlsysaffiliation{iitp}{AI-NLP-ML Research Laboratory, \\Department of Computer Science and Engineering,\\Indian Institute of Technology Patna, Patna, Bihar, India} 
  \mlsyscorrespondingauthor{Parv Agarwal}{parvagarwal9759@gmail.com}
  \mlsyscorrespondingauthor{Asif Ekbal}{asif@iitp.ac.in}
  
\else
  \begin{mlsysauthorlist}
  \mlsysauthor{Anonymous Author(s)}{anon}
  \end{mlsysauthorlist}
  \mlsysaffiliation{anon}{Anonymous Institution (withheld for double-blind review)}
  \mlsyscorrespondingauthor{Anonymous Author(s)}{anon@example.org}
\fi

\mlsyskeywords{GPU clusters, job monitoring, failure diagnosis, observability,
reliability, machine-learning systems}

\vskip 0.3in

\begin{abstract}
GPU training jobs fail often: roughly two in five on large production clusters - yet the operator typically learns of a failure only by reconnecting hours later. Experiment trackers require editing the training script and maintaining a cloud connection; the scheduler's mail hook delivers a single status line with no cause and no logs. GPUAlert is a command-line wrapper that monitors any training command at the process boundary, and with no change to that command, it emails a structured notification on completion carrying a classified failure cause, durable logs, and output artifacts.

The tool is organized around three reliability primitives: a \emph{pre-launch log guarantee} that establishes the durable destination before the child process can crash, \emph{notifier isolation} that makes the wrapper's exit code a pure function of the child's status regardless of whether the email succeeds and a \emph{non-silent artifact budget} that bounds attachment size without ever dropping output silently. We release a labelled corpus of 474 GPU training logs across 15 failure classes, three of which are synthetic, as noted in Section~\ref{sec:eval}, and a reproducible evaluation harness. On the twelve hardware-reproduced classes, the ordered-rule classifier reaches 0.997 macro-F1, against 0.830 for unordered keyword matching and 0.133 for exit-code inspection. Wrapper overhead is a constant $\sim$3\,ms per job; the pre-launch guarantee preserves a log where a shell redirect yields nothing; and across all 15 failure modes the wrapper returns the child's exit code unchanged even when the SMTP relay is unreachable.
\end{abstract}
]
\printAffiliationsAndNotice{}

\section{Introduction}
\label{sec:intro}

The failure of a GPU training job is not a tail event; it is the common case. Large production clusters shed roughly 40\% of their jobs before completion~\citep{hu2024characterization,kokolis2025revisiting,jeon2019analysis,weng2022mlaas}, and infrastructure faults: CUDA errors, ECC events, NVLink and NCCL failures account for the majority of the GPU-hours those failures waste. At thousand-GPU scale, the mean time to the first failure of a run is measured in single-digit hours~\citep{kokolis2025revisiting}. These numbers describe a workload in which something breaks constantly and in which knowing \emph{what} broke, \emph{quickly} translates directly to reclaimed GPU-hours.

The feedback loop available to the person running the job has barely changed. Submit a job, close the laptop, and reconnect hours later to a terminal that shows either a result or a dead process, usually with no context about why it died. The cost is not the failure itself but the latency to noticing it. A job that hits a CUDA out-of-memory error in its third epoch can sit dead for the rest of a twelve-hour reservation while the operator assumes it is still training.

Tools exist, but they do not fit the setting where the gap is widest. Modern experiment trackers are powerful, yet they require importing an SDK into the training script and, in the common configuration, an outbound connection to a hosted backend. Neither assumption holds for an inherited script, a quick one-off run, a shared cluster with restrictive egress, or an air-gapped HPC partition. At the other extreme, the scheduler's notification hook requires no code change but carries almost no information: one line saying ``the job ended``, with no logs, no cause, and no artifacts. Between ``instrument everything and phone a cloud'' and ``a single status line'' there is a wide unserved middle, and it is where most day-to-day training happens.

This paper presents \textbf{GPUAlert}, a tool built for that middle. GPUAlert wraps a training command and monitors it at the \emph{process boundary}: it observes the child process's standard streams and exit status rather than instrumenting the Python program inside\footnote{Process-boundary operation means unmodified commands: no source edits, no imports.}
\\When the job ends for any reason, GPUAlert classifies the failure from the captured output, gathers output artifacts under a size budget, and sends one structured email containing the cause, a remediation hint, the logs, and the artifacts. Installation is a single package install; the common case is a one-line prefix on an existing command.\footnote{Source, corpus and harness:
\ifpreprint
\url{https://github.com/Parv-01/gpualert} and \url{https://github.com/Parv-01/gpualert-eval}.
\else
\url{https://anonymous.4open.science/r/gpualert/} and \url{https://anonymous.4open.science/r/gpualert-eval/}.
\fi}

For a process-boundary wrapper to be trusted in front of every job, it must fail gracefully. A monitor that drops logs during a crash or corrupts exit codes when a mail server times out is worse than no monitor at all in an automated pipeline. \textit{GPUAlert} solves this through three core reliability primitives: a \emph{pre-launch log guarantee} that secures the output destination before execution; \emph{notifier isolation} to ensure the final exit code mirrors the child process; and an \emph{artifact budget} that prevents silent data loss. While these borrow established concepts: write-ahead logging, the bulkhead pattern, and graceful degradation-their integration and evaluation at the GPU process boundary forms our main contribution.

\paragraph{Contributions.}
\begin{itemize}
\itemsep2pt
\item A process-boundary monitoring design for GPU training jobs, requiring zero instrumentation of the workload and organized around three reliability primitives stated precisely in Section~\ref{sec:design} and evaluated in Section~\ref{sec:eval}. 
\item A priority-ordered failure classifier covering 15 GPU and Python failure modes, each paired with a remediation hint delivered in the notification email.
\item An open, labelled corpus of 474 GPU training logs and a reproducible evaluation harness released as artifacts.
\item An empirical evaluation: 0.997 macro-F1 over twelve hardware-reproduced classes, $\sim$ 3\,ms constant per-job overhead, log durability where shell redirection fails and exit-code preservation across all 15 modes when the SMTP relay is unreachable.
\end{itemize}

\section{Background and Related Work}
\label{sec:related}

\paragraph{Why job failures dominate GPU operations.}
\citet{hu2024characterization} characterize six months of an LLM development datacenter and report that a large fraction of jobs fail before completion, with infrastructure faults responsible for the bulk of wasted GPU time and failures concentrated early in a run. \citet{kokolis2025revisiting} revisit failures across roughly 150 million A100 GPU-hours at Meta and find a mean time to first failure of approximately 7.9 hours at the 1024-GPU scale. \citet{jeon2019analysis} document the same qualitative pattern in the Microsoft Philly trace and \citet{weng2022mlaas}
report comparable churn in Alibaba's multi-tenant clusters. Together these establish two facts this work depends on: failures are frequent, and the dominant cost is the GPU-time lost between a failure and its discovery. GPUAlert attacks the second

\paragraph{Experiment trackers.}
Weights \& Biases~\citep{biewald2020wandb}, MLflow~\citep{zaharia2018mlflow} and Comet ML~\citep{cometml} record metrics, artifacts and run metadata with rich visualization. They are the right tool once a pipeline is owned and instrumented, but they require importing an SDK into the training program and, in the common configuration, an outbound connection to a hosted backend. \citet{shankar2022operationalizing}, interviewing ML practitioners, find this
instrumentation and connectivity cost to be a recurring source of friction, especially on inherited code. GPUAlert is complementary rather than competing: it needs no import, no account, and no connectivity beyond an SMTP relay.

\paragraph{Daemon-level observability.}
Log aggregation stacks (Elasticsearch, Loki) and GPU hardware monitors such as NVIDIA DCGM~\citep{dcgm} answer cluster-level and device-level questions and are rightly the concern of site administrators. They require daemon deployment and elevated access, and they present an aggregate view across jobs rather than per-job, per-user notification. GPUAlert is orthogonal: it runs as an unprivileged user-space process and scopes its output to the one job the user launched.

\paragraph{Decorator-based notification.}
knockknock~\citep{knockknock} attaches a Python decorator to a training function and can notify across many channels. It is well suited to a script the user controls and wants breadth of delivery for, but it requires editing the code, does not classify the failure cause, and provides no guarantee that a log survives a crash.

\paragraph{Scheduler hooks.}
Slurm's \texttt{-{}-mail-type}~\citep{yoo2003slurm} fires on job state transitions but carries no log content and no diagnosis; it is precisely the exit-code-only baseline in our evaluation.

\paragraph{Checkpointing.}
CheckFreq~\citep{mohan2021checkfreq} and Check-N-Run~\citep{eisenman2022checknrun} reduce the work lost to a failure by frequent, cheap checkpointing. They are orthogonal: checkpointing limits the damage of a failure; GPUAlert reduces the time to learn of and diagnose one. The two compose naturally.

\begin{figure*}[t]
\centering
\resizebox{\textwidth}{!}{%
\begin{tikzpicture}[
  font=\scriptsize,
  box/.style={rectangle, rounded corners=2pt, draw=black!55, fill=blue!4,
              align=center, minimum height=0.85cm, text width=1.85cm, inner sep=2pt},
  prim/.style={rectangle, rounded corners=2pt, draw=orange!70, fill=orange!8,
               align=center, font=\scriptsize\itshape, text width=2.3cm,
               minimum height=0.7cm, inner sep=2pt},
  arr/.style={-Stealth, thick, draw=black!55},
  darr/.style={->, dashed, draw=orange!75, thick},
]
\node[box] (cli) {\texttt{cli}\\invoke + parse};
\node[box, right=0.5cm of cli] (log) {\texttt{log\_manager}\\touch 3 logs};
\node[box, right=0.5cm of log] (pop) {\texttt{launcher}\\\texttt{Popen} + 2 stream threads};
\node[box, right=0.5cm of pop] (wait) {\texttt{wait()}\\/ \texttt{kill} on timeout};
\node[box, right=0.5cm of wait] (parse) {\texttt{parse\_errors}\\classify + metrics};
\node[box, right=0.5cm of parse] (art) {\texttt{artifacts}\\scan + budget};
\node[box, right=0.5cm of art] (notif) {\texttt{notifier}\\build + SMTP};
\node[box, right=0.5cm of notif] (exit) {\texttt{exit(}\\\texttt{child\_code)}};
\draw[arr](cli)--(log);\draw[arr](log)--(pop);\draw[arr](pop)--(wait);
\draw[arr](wait)--(parse);\draw[arr](parse)--(art);\draw[arr](art)--(notif);
\draw[arr](notif)--(exit);
\node[prim, below=0.7cm of log] (p1) {P1: pre-launch log guarantee};
\node[prim, below=0.7cm of art] (p3) {P3: artifact budget};
\node[prim, below=0.7cm of exit, xshift=-0.4cm] (p2) {P2: notifier isolation};
\draw[darr](p1)--(log);
\draw[darr](p3)--(art);
\draw[darr](p2.north)--(notif.south);
\draw[darr](p2.north east)--(exit.south);
\end{tikzpicture}}
\caption{GPUAlert execution flow. The wrapped command runs verbatim as a child process; the three reliability primitives (P1--P3, dashed) sit at the stages where a naive wrapper would lose data or corrupt the exit code.}
\label{fig:arch}
\vspace{-10pt}
\end{figure*}
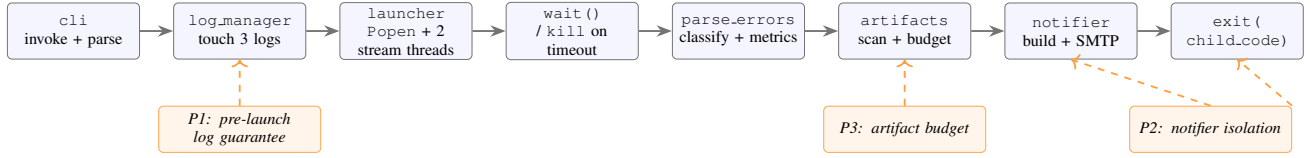
\section{Design and Implementation}
\label{sec:design}

GPUAlert is a Python command-line tool. The core invocation prefixes an existing command as in Listing~\ref{lst:run}. Everything after \texttt{-{}-} is the wrapped command run verbatim. The wrapper launches it as a child process, streams its standard output and error to per-job log files in real time, waits for it (enforcing an optional timeout), classifies the outcome from the captured output, collects output artifacts under a size budget, and sends one structured email. The exit code returned to the caller is always the child's. Figure~\ref{fig:arch} shows the
execution flow and where the three reliability primitives sit.

\begin{lstlisting}[language=bash,caption={Wrapping a training command. The wrapped
command runs verbatim; GPUAlert adds monitoring around it without touching it.},
label={lst:run}]
gpualert run -- python train.py --epochs 20
\end{lstlisting}
\vspace{-10pt}
\subsection{Design constraint: no change to the wrapped command}
The primary constraint governing \texttt{GPUAlert} is strict execution equivalence: the wrapped command must run exactly as it would in isolation. This operational boundary precludes any modifications to the underlying source code or runtime process injection. Consequently, the tool remains entirely environment-agnostic, making it immediately deployable on legacy codebases, within pre-built containerized environments, or on managed cluster nodes where 
administrative privileges are restricted. By enforcing this zero-instrumentation invariant, all downstream architecture decisions rely strictly on black-box observation of the process from the outside, rather than introspection from within.

\subsection{P1: Pre-launch log guarantee}
To guarantee data durability, the \texttt{log\_manager} initializes and allocates the three core log artifacts (\texttt{stdout.log}, \texttt{stderr.log} and \texttt{combined.log}) on disk \emph{prior} to subprocess instantiation, as detailed in Listing~\ref{lst:prelog}.\\
\pagebreak[4]
\begin{lstlisting}[language=Python,caption={Pre-launch guarantee: the log files
exist on disk before the subprocess starts. If the child segfaults on its first
instruction, the destination is already durable.},label={lst:prelog}]
_LABEL = "user/pre-launch"  # internal trace tag
for log_file in (stdout_log, stderr_log, combined_log):
    log_file.parent.mkdir(parents=True, exist_ok=True)
    log_file.touch(exist_ok=True)
proc = subprocess.Popen(cmd, stdout=out_fh, stderr=err_fh)
\end{lstlisting}

\noindent The analogy is write-ahead logging: the durable destination is committed before the operation that might fail. If the child segfaults on its first instruction or the executable does not exist and \texttt{Popen} itself fails, the log files are present and will be attached to the notification. The guarantee is deliberately narrow: it ensures the destination \emph{exists}, not that unflushed buffers are recovered; but as Section~\ref{sec:eval} shows, that narrow guarantee is exactly the difference between a usable diagnostic and an empty directory when a command never starts.

\subsection{Failure classification}
Upon process termination, \texttt{parse\_errors} evaluates the combined log against an ordered list of regular expressions using a first-match-wins strategy. To prevent generic catch-alls from overshadowing precise diagnostics, rule prioritization dictates that highly specific patterns take precedence over broader ones. For instance, explicit \texttt{AssertionError} and \texttt{RuntimeError} rules are assigned higher priorities than the generic Python \texttt{Traceback} filter. This hierarchy was established a priori, before log evaluation. 

Table~\ref{tab:taxonomy} outlines the final 15 failure modes in priority order alongside their corresponding email remediation hints. The classifier is intentionally lightweight, relying on text-based regular expressions rather than a trained model. This design ensures the diagnostics remain dependency-free, auditable, and fast, while maintaining high accuracy across the targeted failure surface (Section~\ref{sec:eval}).

\begin{table}[t]
\vspace{-10pt}
\caption{Failure taxonomy in priority order; first match wins. Rows are ordered from most-specific (top) to most-general (bottom). Each rule emits the remediation hint shown in the notification body.}
\label{tab:taxonomy}
\vskip 0.1in
\centering\scriptsize\setlength{\tabcolsep}{3pt}\renewcommand{\arraystretch}{1.2}
\begin{tabular}{@{}rp{1.5cm}p{2.75cm}p{2.55cm}@{}}
\toprule
\textbf{Pri.} & \textbf{Label} & \textbf{Pattern (excerpt)} & \textbf{Hint} \\
\midrule
1  & CUDA OOM        & \texttt{CUDA out of memory}        & Lower batch / grad.\ ckpt \\
2  & NCCL error      & \texttt{NCCL error}                & Check fabric; \texttt{NCCL\_DEBUG} \\
3  & CUDA runtime    & \texttt{RuntimeError: CUDA}        & Check \texttt{nvidia-smi}; GPU fault \\
4  & System OOM      & \texttt{MemoryError}               & Reduce RAM / add swap \\
5  & Segfault        & \texttt{Segmentation fault}        & Check native/CUDA libs \\
6  & File missing    & \texttt{FileNotFoundError}         & Check data paths \\
7  & Permission      & \texttt{PermissionError}           & Check file permissions \\
8  & Missing module  & \texttt{ModuleNotFoundError}       & \texttt{pip install -r reqs} \\
9  & Div by zero     & \texttt{ZeroDivisionError}         & Check denominators \\
10 & Device mismatch & \texttt{tensors.*same device}      & Align tensor devices \\
11 & NaN loss        & \texttt{loss.*nan}                 & Check LR / clipping \\
12 & OOM-killer      & \texttt{Out of memory: Kill}       & Request more RAM \\
13 & Assertion       & \texttt{AssertionError}            & Check shapes / assumptions \\
14 & RuntimeError    & \texttt{RuntimeError}              & Check \texttt{stderr.log} \\
15 & Traceback       & \texttt{Traceback (most\ldots)}    & See attached log \\
\bottomrule
\end{tabular}
\vspace{-15pt}
\end{table}

\subsection{P2: Notifier isolation}
The email send is wrapped so that any exception---DNS failure, refused connection, authentication error, timeout--- is caught, recorded to a local log, and not re-raised, as in Listing~\ref{lst:isolation}.

\begin{lstlisting}[language=Python,caption={Notifier isolation: a failed send is caught and recorded locally, so the wrapper's exit code remains a pure function of the child's status.},label={lst:isolation}]
try:
    smtp.send_message(msg)
except Exception:        # bulkhead: never propagate
    _log_local_failure() # record to ~/.gpualert/send_errors.log
# control always reaches:
sys.exit(job_exit_code)

\end{lstlisting}

\noindent This is the bulkhead pattern: a failure in the notification compartment cannot reach the job-status compartment. The wrapper's exit code is a pure function of the child's exit code. This matters wherever an exit code is load-bearing: Makefile rules, CI steps, and Slurm job arrays all branch on it, and a monitor that turned a successful job into a failure because the SMTP relay was unreachable would be actively harmful. Section~\ref{sec:eval} verifies the property holds across every failure mode.

\subsection{P3: Non-silent artifact budget}
\texttt{artifacts} scans the working directory for output files and attaches them up to a configurable per-file byte cap (25\,MB by default). A file that exceeds the cap is not silently dropped; it is listed under a \texttt{Skipped:} heading in the email body with its name and size, so the operator knows the artifact exists and where to retrieve it. The combined log is always attached on failure regardless of the budget. The principle is graceful, visible degradation: bounded behaviour under large outputs, with no silent data loss.

\subsection{Scheduler integration, configuration and security}
For batch environments, GPUAlert can attach to an already-submitted job rather than wrapping a launch: \texttt{gpualert slurm <jobid>} polls \texttt{sacct} at a configurable interval and on a terminal state, routes through the same notification pipeline. No daemon runs on the compute node; the poller runs wherever the user has \texttt{sacct} access. This mode is not evaluated in the current work and should be treated as experimental; Section~\ref{sec:discussion} lists it as a planned extension. 

Configuration lives in a single TOML file under the user's home directory, written with \texttt{0600} permissions. SMTP credentials are stored in plaintext, which is no worse than an SSH key or a \texttt{.netrc} file under the same permission model, but operators in shared environments should use an app-specific password or a dedicated relay account. The interactive setup validates the SMTP host before accepting the configuration, so a malformed entry fails at setup rather than silently at the first real job.

\section{Evaluation}
\label{sec:eval}

Our evaluation answers four questions, one per reliability claim plus the classifier: (Q1) how accurately does the rule classifier identify failure causes, against realistic baselines? (Q2) Does the pre-launch log guarantee preserve a diagnostic where simpler methods do not? (Q3) What does the wrapper cost? (Q4) Is the exit code truly isolated from mail failures? We add a fifth check on the artifact budget.

\paragraph{Setup.}
All experiments ran on a single \textbf{NVIDIA Tesla V100-PCIE-32GB} (driver 460.91.03, PyTorch 2.7.1+cu118~\citep{paszke2019pytorch}, Python 3.13). The evaluation is deliberately bounded to one hardware configuration; the implications of this are discussed in Section~\ref{sec:discussion}. Every number below is
reproducible from a single \texttt{make all} target~\citep{gpualert_eval}.

\paragraph{Corpus.}
We assembled \textbf{474 labelled GPU training logs} across 15 failure classes. 360 were captured on the real V100 by injecting each fault into a training loop (12 classes). Three classes---\texttt{cuda\_oom}, \texttt{nccl} and
\texttt{oom\_killer}---could not be faithfully reproduced on this single, old-driver, unprivileged node (an NVML symbol incompatibility blocked a true CUDA OOM, NCCL requires $\geq$2 ranks and the OOM-killer requires a memory-capped
cgroup). These three classes use recorded reference signatures tagged \texttt{source: synthetic} in the released corpus. \textbf{Results on these three classes should not be read as evidence of real-world detection}; the conservative figure throughout is the 0.997 on the twelve hardware-reproduced classes. A further 24 logs were drawn from public issue trackers as a held-out ``wild'' set the classifier never saw during rule development.
\vspace{-15pt}
\paragraph{Baselines.}
\texttt{exitcode} classifies on the process exit code alone---what Slurm's mail hook effectively provides. \texttt{traceback} parses the Python exception type from stderr. \texttt{grep} applies a naive first-match keyword search with no priority ordering over the same 15 pattern strings.

\subsection{Q1: Classification accuracy}
Table~\ref{tab:clf} reports macro-F1 with 95\% bootstrap confidence intervals over the full 474-log corpus. GPUAlert reaches \textbf{0.997 macro-F1} on the twelve hardware-reproduced classes; including the three synthetic classes pushes this to 0.998. The ordered rules dominate every baseline: unordered keyword matching trails by 17 points, exception-type parsing by more than 40, and exit-code inspection -- the de facto status quo for unmodified jobs -- is barely above chance because most distinct failure modes share a non-zero exit code.

\begin{table}[!ht]
\vspace{-12pt}
\caption{Classification on 474 logs, 15 classes. \textit{Real-12} excludes the three synthetic classes; those three classes have no real-hardware basis, and the Real-12 figure is the conservative claim.}
\label{tab:clf}
\vskip 0.1in
\centering\small
\begin{tabular}{@{}lcccc@{}}
\toprule
\textbf{Classifier} & \textbf{Macro-F1} & \textbf{95\% CI} & \textbf{Real-12} & \textbf{Acc.} \\
\midrule
\textbf{GPUAlert} & \textbf{0.998} & [0.993, 1.000] & \textbf{0.997} & \textbf{0.998} \\
grep              & 0.830 & [0.822, 0.838] & ---            & 0.859 \\
traceback         & 0.559 & [0.552, 0.565] & ---            & 0.601 \\
exitcode          & 0.133 & [0.133, 0.133] & ---            & 0.131 \\
\bottomrule
\end{tabular}
\vspace{-10pt}
\end{table}

The gap is statistically unambiguous. McNemar's exact test (GPUAlert vs.\ grep): of 474 logs, both are correct on 406, GPUAlert alone on 67, grep alone on 1; $p = 4.68\times10^{-19}$. On the 24 held-out wild logs, GPUAlert reaches \textbf{95.8\%} accuracy (23/24; Wilson 95\% CI [0.79, 1.00]) versus 70.8\% for grep and 62.5\% for traceback parsing. The wild-set sample is small, and the confidence interval reflects that; the result is consistent with the main corpus
finding rather than a strong independent claim.

\begin{figure}[t]
\centering
\includegraphics[width=0.9\columnwidth, trim=0pt 50pt 0pt 0pt, clip]{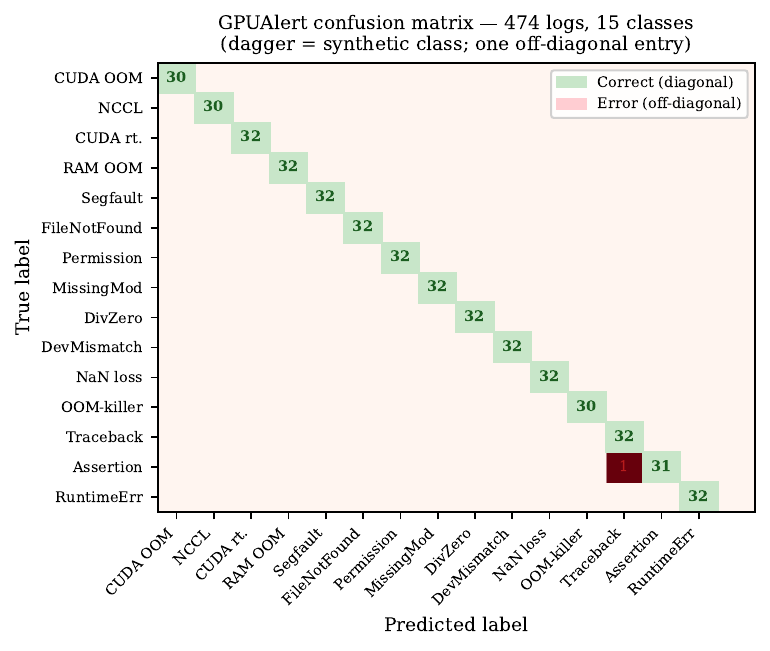}
\caption{GPUAlert confusion matrix over the 474-log corpus. The single off-diagonal entry is one \texttt{assertion} log classified as \texttt{traceback}, the boundary case motivating the priority ordering in Table~\ref{tab:taxonomy}.}
\label{fig:confusion}
\vspace{-10pt}
\end{figure}

Per class (Figure~\ref{fig:confusion}), 13 of 15 classes reach F1 $= 1.000$. The single off-diagonal entry is one \texttt{assertion} log that surfaces only a bare \texttt{Traceback} line and lands in the traceback class---the boundary case that motivates placing the generic Traceback rule at lowest priority.

\subsection{Q2: Log durability}
We triggered four fault types: \texttt{python\_exception}, \texttt{segfault}, \texttt{SIGKILL}, and \texttt{exec\_failure} (the executable does not exist), and checked whether GPUAlert, a shell redirect (\texttt{> log 2>\&1}), and \texttt{nohup} each left a usable log. For the first three faults, where the process starts and flushes before dying, all three methods leave a non-empty log in 20/20 trials: \emph{GPUAlert is not uniquely better here, and we say so explicitly}. The separation is on \texttt{exec\_failure}: GPUAlert leaves a non-empty log in 20/20 trials because the destination was touched before the launch attempt, while the shell redirect leaves nothing in 20/20 (Fisher exact, $p = 1.45\times10^{-11}$). The guarantee is about the durable destination existing, and it converts a silent ``empty directory'' failure into a diagnosable one.

\subsection{Q3: Wrapper overhead}

Table~\ref{tab:overhead} reports wall-clock overhead over 30 trials (warm-ups discarded, Welch's $t$). The penalty is $+3.78$\,ms on a no-op and $+2.61$\,ms on a short Python script.

\begin{table}[!ht]
\vspace{-10pt}
\centering\small
\caption{Wrapper overhead (30 trials each, Welch's $t$). Cost is constant per job, not proportional to job length; on any job running longer than a few seconds the penalty is negligible.}
\label{tab:overhead}
\vskip 0.15in
\begin{tabular}{@{}lrrrr@{}}
\toprule
\textbf{Workload} & \textbf{Bare} & \textbf{Wrapped} & \textbf{Overhead} & \textbf{p} \\
\midrule
noop      & 36.32\,ms & 40.10\,ms & $+3.78$\,ms & $8.5\times10^{-4}$ \\
short\_py & 42.65\,ms & 45.26\,ms & $+2.61$\,ms & $2.0\times10^{-5}$ \\
\bottomrule
\end{tabular}
\vspace{-10pt}
\end{table}

The overhead is constant per job classification, and the mail send happens once, after the child exits, not on any per-step path, so it becomes invisible against any job measured in minutes or hours. We do not claim zero overhead; we claim a fixed $\mathcal{O}(1)$ cost of a few milliseconds.

\subsection{Q4: Notifier isolation}
The SMTP relay was disabled by pointing \texttt{SMTP\_HOST} to a non-listening port, producing a refused-connection exception on every send attempt. We then ran all 15 failure modes through the wrapper. In every case the wrapper returned the child's exit code unchanged, and no exception escaped the notifier: 15/15. The exit code is a pure function of the child status, as the bulkhead design intends.

\subsection{Artifact budget}
We ran seven artifact sizes against the default 25\,MB cap: 1\,KB, 100\,KB, 1\,MB, 10\,MB (all attached), 30\,MB, 60\,MB and 1\,GB (all listed under \texttt{Skipped:} with their name and size). The combined log was delivered on failure in every case. Behaviour degrades predictably as outputs grow and nothing is dropped silently.

\subsection{End-to-end notifications}
Figure~\ref{fig:email} shows the two notification shapes an operator actually receives: a success report carrying duration, exit code and extracted final metrics, and a failure report carrying the classified cause, the remediation hint and the attached logs. This arrives within seconds of the job ending, without the operator having reconnected.

\begin{figure}[h]
\centering
\resizebox{0.90\columnwidth}{!}{%
\begin{tikzpicture}[font=\scriptsize\ttfamily, node distance=0.3cm]
\node[draw=black!45, rounded corners=2pt, fill=green!4, text width=7.2cm, align=left, inner sep=4pt] (succ) {
  \textbf{[GPUAlert] SUCCESS train.py}\\[1pt]
  \scriptsize\rmfamily From: gpualert@host\quad To: you@example.org\\
  \hrulefill\\
  Status: SUCCESS\\
  Command: python train.py --epochs 20\\
  Duration: 6h 02m 14s \quad Exit: 0\\
  Metrics: val\_acc 0.9312 \quad loss 0.2104\\
  Attached: combined.log, checkpoint listed
};
\node[draw=black!45, rounded corners=2pt, fill=red!4, text width=7.2cm, align=left, inner sep=4pt, below=of succ] (fail) {
  \textbf{[GPUAlert] FAILED train.py}\\[1pt]
  \scriptsize\rmfamily From: gpualert@host\quad To: you@example.org\\
  \hrulefill\\
  Status : FAILED \quad Exit: -2\\
  Command: python train.py --epochs 20\\
  Cause: CUDA out of memory (epoch 3)\\
  Hint: lower batch size or grad.\ checkpointing\\
  Attached: stdout.log, stderr.log, combined.log\\
  Skipped: ckpt\_epoch3.pt (1.2\,GB, over 25\,MB cap)
};
\end{tikzpicture}}
\caption{Anonymized mockups of the two notification shapes. Success carries metrics and artifacts; failure carries the classified cause, a remediation hint, and durable logs, with oversized artifacts listed rather than dropped or emailed.}
\label{fig:email}
\vspace{-10pt}
\end{figure}

\section{Discussion and Positioning}
\label{sec:discussion}

Table~\ref{tab:comparison} positions GPUAlert against the tools an operator would otherwise reach for. GPUAlert does not replace experiment trackers on well-instrumented pipelines, and it has nothing to offer operators who need notification across a dozen chat channels: knockknock handles that better. What it occupies is the slot those tools leave empty: zero-code instrumentation, a classified per-job cause, durable logs and exit-code safety, all at the process boundary and all runnable on a cluster with no outbound web access. Each cell where GPUAlert differs from the alternatives reflects a property directly measured in Section~\ref{sec:eval}.

\begin{table}[h]
\vspace{-10pt}
\caption{Feature comparison. GPUAlert is the only entry combining zero instrumentation with per-job failure classification, durable logs, and exit-code isolation.}
\label{tab:comparison}
\vskip 0.1in
\centering\small\setlength{\tabcolsep}{3.5pt}\renewcommand{\arraystretch}{1.15}
\resizebox{\columnwidth}{!}{%
\begin{tabular}{@{}lccccc@{}}
\toprule
\textbf{Property} & \textbf{GPUAlert} & \textbf{W\&B} & \textbf{knock-} &
\textbf{Slurm} & \textbf{DCGM} \\
                  &                   &               & \textbf{knock}  &
\textbf{mail}  &               \\
\midrule
Zero code change          & \cmark & \xmark  & \xmark  & \cmark & \cmark \\
15-way cause class.\      & \cmark & \xmark  & \xmark  & \xmark & hw.\ only \\
Pre-launch log guar.\     & \cmark & \xmark  & \xmark  & \xmark & n/a \\
Log/artifact delivery     & \cmark & part.\  & \xmark  & \xmark & \xmark \\
Air-gap / no cloud        & \cmark & \xmark  & \xmark  & \cmark & \cmark \\
Notifier isolation        & \cmark & n/a     & \xmark  & n/a    & n/a \\
Channels                  & email  & web     & 12+     & email  & n/a \\
\bottomrule
\end{tabular}%
}
\vspace{-10pt}
\end{table}

\noindent DCGM ``hw.\ only'' denotes that it surfaces hardware events---ECC errors, XID faults, thermal throttling---at the device level but does not classify why a particular training job failed at the Python or framework level.

\paragraph{Limitations.}
The evaluation is bounded to a single V100 running driver 460.91.03; all 360 hardware-reproduced logs originate from that one machine. The 24 held-out logs drawn from public issue trackers are encouraging, but evaluating across A100, H100, and consumer RTX hardware over multiple driver generations is necessary before claiming broad generalization. Three of the 15 failure classes remain synthetic by necessity---\texttt{cuda\_oom}, \texttt{nccl}, and \texttt{oom\_killer} could not be faithfully reproduced on this unprivileged, single-GPU node and results on those
three reflect only that the classifier matches patterns it was designed to match.

The regex ruleset covers the failure surface that production-cluster studies report most frequently; a framework or runtime emitting non-standard error strings will fall through to the generic \texttt{Traceback} rule, which is informative but not specific. New failure modes are cheap to add---one rule, one corpus entry---but each addition requires real hardware evidence to avoid carrying the same limitation the current synthetic classes do.

The more fundamental gap is multi-node distributed training. GPUAlert monitors a single process boundary and is well-suited to single-GPU and single-node jobs. Multi-node runs surface qualitatively different failures---rank hangs, node dropout mid-run, NCCL rendezvous timeouts, silent gradient divergence across ranks---that may produce no output on the monitored process before it times out or is killed. Closing this gap requires either a coordinator-side wrapper that aggregates termination signals from all participant ranks or integration with the process
launcher (\texttt{torchrun}, \texttt{mpirun}) that already holds visibility into the collective. The cluster studies motivating this work~\citep{kokolis2025revisiting, hu2024characterization} operate at exactly this scale; extending GPUAlert there is the work we consider most pressing.

\paragraph{Environmental considerations.}
Compute infrastructure for large-scale ML training carries a well-documented energy and carbon cost~\citep{hu2024characterization,kokolis2025revisiting}. Failed jobs that go undetected represent a particularly wasteful slice of that budget. A GPU does not cease drawing power when its training process crashes: the allocation stays active, cooling systems continue running, and the device idles at non-trivial wattage until a human notices and releases the reservation. At the failure rates the cluster studies report: roughly 40\% of jobs failing before completion, with mean times to first failure measured in single-digit hours against reservations that often span twelve or more; the expected dead-running time per failed job accumulates quickly.

GPUAlert does not prevent failures, but it eliminates the detection latency. A notification arriving within seconds of job termination lets the operator release the allocation, correct the fault, and resubmit--rather than letting the reservation exhaust its wall-clock budget on a process that stopped making progress hours earlier. At cluster scale, the aggregate effect is a measurable reduction in wasted GPU-hours, which translates into avoided energy expenditure and a lower operational carbon footprint. This is not the primary design motivation but it is a consequence worth stating: prompt failure notification is not only operationally useful, it is also a concrete- if modest, step toward more responsible use of scarce compute resources.

\paragraph{Future work.}
Several directions follow naturally from the current design, roughly in order of expected impact.

\textit{Multi-node coverage.} A coordinator-side extension that wraps the launcher process and monitors the collective termination status of all participant ranks would bring the same notification and classification pipeline to distributed training without modifying the training script. The core difficulty is partial-rank failure: some processes may exit with a meaningful code while others hang, and the wrapper must aggregate across potentially heterogeneous signatures before committing to a single classified cause.

\textit{Live anomaly detection.} The current design observes at exit; it is blind to a job that is running but making no progress: a stalled loss curve, a GPU utilization collapse to near-zero, or a training loop cycling on corrupted data. Periodic sampling of the captured output stream during execution, paired with lightweight pattern detection on metric lines, could surface these pathologies before the reservation expires rather than after it, further closing the gap between fault onset and operator awareness.

\textit{Delivery channels and scheduler integration.} Slack and Microsoft Teams webhooks, PagerDuty integrations, and generic HTTP endpoints would cover the notification preferences of most teams without touching the classification or monitoring core. On the scheduler side, Slurm epilog integration - triggering the classifier and notifier as an epilog script rather than through polling - would eliminate sacct latency and compose naturally with multi-node job steps.

\textit{Corpus expansion and learned classification.} The current 474-log corpus spans one hardware generation and 15 failure classes. A community-contributed extension covering A100 and H100 hardware, JAX and TensorFlow runtimes, and less common failure modes (NVLink errors, PCIe bandwidth saturation, driver version mismatches) would validate the existing rules and expose the cases where regex matching is insufficient. Once the failure surface is broad enough and the class boundaries are clear enough, a lightweight learned classifier---trained on the open corpus and shipped alongside the rules---could handle non-standard error strings that currently fall through to the generic \texttt{Traceback} rule.

\section{Conclusion}
\label{sec:conclusion}

A GPU cluster running at production scale fails roughly two jobs in five before they complete. The direct cost---lost progress, wasted reservation---is well-documented. The indirect cost, less often measured, is the latency between when a job stops and when the operator learns of it. A job that crashes in its third epoch and sits unreported for four hours has not merely wasted four hours of GPU time; it has also delayed whatever decision or correction follows. GPUAlert is built on the premise that closing this notification gap reliably, without touching the training
script is a problem worth solving precisely.

The design is deliberately conservative. Rather than asking for instrumentation rights, a cloud connection or a deployed daemon, the tool wraps the one interface any command exposes: its process boundary. Three correctness properties are treated as non-negotiable---the log must exist before the child can crash, the exit code must reflect the job, not the mail server, and artifacts must never disappear without acknowledgement. None of these properties is individually novel; write-ahead logging, the bulkhead pattern, and graceful degradation are well-understood abstractions. The contribution is their combination at the process boundary, a 474-log corpus and reproducible harness that allows the properties to be verified rather than assumed, and a failure classifier that reaches 0.997 macro-F1 on hardware-reproduced fault classes while adding a fixed three-millisecond overhead per job. Faster detection also carries an environmental consequence: every hour shaved from the gap between failure and discovery is an hour of allocated but idle GPU compute---and its associated energy draw---returned to productive use.

The open questions are honest ones. The evidence is strongest for single-GPU, single-driver deployments; the operational need is greatest for multi-node distributed training, where the failure surface spans rank boundaries and the process interface that GPUAlert currently monitors carries only a fraction of the relevant signal. Bridging that gap---monitoring a collective of processes across nodes, classifying failures that no single process fully observes, doing so without modifying the launcher---is a harder problem than the one addressed here. That the current design handles the single-node case correctly, completely and reproducibly is offered as a foundation for that work, not as a substitute for it.

\ifpreprint
\section*{Acknowledgements}
The core design, implementation and evaluation of GPUAlert are the authors' own work. Large language models were used to assist with code formatting, boilerplate unit tests and prose editing; all experimental design, scientific claims and interpretation of results are the authors' own.\\
The authors gratefully acknowledge the project ``Centre of Indian Language Data (COIL-D)" under the flagship mission of Bhashini, funded by MeitY, Government of India, for the financial grant that enabled the successful conduct of this research.
\fi

\bibliographystyle{mlsys2025}
\bibliography{references}

\begin{thebibliography}{15}
\providecommand{\natexlab}[1]{#1}
\providecommand{\url}[1]{\texttt{#1}}
\expandafter\ifx\csname urlstyle\endcsname\relax
  \providecommand{\doi}[1]{doi: #1}\else
  \providecommand{\doi}{doi: \begingroup \urlstyle{rm}\Url}\fi

\bibitem[Biewald(2020)]{biewald2020wandb}
Biewald, L.
\newblock Experiment tracking with weights and biases, 2020.
\newblock URL \url{https://wandb.ai/site}.
\newblock Software.

\bibitem[{Comet ML Inc.}(2021)]{cometml}
{Comet ML Inc.}
\newblock {Comet ML}: A meta machine learning platform, 2021.
\newblock URL \url{https://www.comet.com/site/}.
\newblock Software.

\bibitem[Eisenman et~al.(2022)Eisenman, Matam, Ingram, et~al.]{eisenman2022checknrun}
Eisenman, A., Matam, K.~K., Ingram, S., et~al.
\newblock {Check-N-Run}: A checkpointing system for training deep learning recommendation models.
\newblock In \emph{19th USENIX Symposium on Networked Systems Design and Implementation (NSDI 22)}, pp.\  929--943, 2022.

\bibitem[Hu et~al.(2024)Hu, Ye, Wang, Wang, Zhang, Chen, Sun, Lin, Wang, Luo, et~al.]{hu2024characterization}
Hu, Q., Ye, Z., Wang, Z., Wang, G., Zhang, M., Chen, Q., Sun, P., Lin, D., Wang, X., Luo, Y., et~al.
\newblock Characterization of large language model development in the datacenter.
\newblock In \emph{21st USENIX Symposium on Networked Systems Design and Implementation (NSDI 24)}, pp.\  709--729, 2024.

\bibitem[{Hugging Face}(2019)]{knockknock}
{Hugging Face}.
\newblock knockknock: Get notified when your training ends, 2019.
\newblock URL \url{https://github.com/huggingface/knockknock}.
\newblock Software.

\bibitem[Jeon et~al.(2019)Jeon, Venkataraman, Phanishayee, Qian, Xiao, and Yang]{jeon2019analysis}
Jeon, M., Venkataraman, S., Phanishayee, A., Qian, J., Xiao, W., and Yang, F.
\newblock Analysis of large-scale multi-tenant {GPU} clusters for {DNN} training workloads.
\newblock In \emph{2019 USENIX Annual Technical Conference (USENIX ATC 19)}, pp.\  947--960, 2019.

\bibitem[Kokolis et~al.(2025)Kokolis, Kuchnik, Hoffman, et~al.]{kokolis2025revisiting}
Kokolis, A., Kuchnik, M., Hoffman, J., et~al.
\newblock Revisiting reliability in large-scale machine learning research clusters.
\newblock In \emph{2025 IEEE International Symposium on High Performance Computer Architecture (HPCA)}, pp.\  1259--1274, 2025.
\newblock \doi{10.1109/HPCA61900.2025.00096}.

\bibitem[Mohan et~al.(2021)Mohan, Phanishayee, and Chidambaram]{mohan2021checkfreq}
Mohan, J., Phanishayee, A., and Chidambaram, V.
\newblock {CheckFreq}: Frequent, fine-grained {DNN} checkpointing.
\newblock In \emph{19th USENIX Conference on File and Storage Technologies (FAST 21)}, pp.\  203--216, 2021.

\bibitem[{NVIDIA Corporation}(2024)]{dcgm}
{NVIDIA Corporation}.
\newblock {NVIDIA} data center {GPU} manager ({DCGM}), 2024.
\newblock URL \url{https://developer.nvidia.com/dcgm}.
\newblock Software.

\bibitem[Parv(2026)]{gpualert_eval}
Parv, A.
\newblock {gpualert-eval}: Evaluation corpus and harness, 2026.
\newblock URL \url{https://github.com/Parv-01/gpualert-eval/}.
\newblock Double-blind artifact(474 labelled GPU training logs, 15 failure classes, five experiments).

\bibitem[Paszke et~al.(2019)Paszke, Gross, Massa, Lerer, Bradbury, Chanan, Killeen, Lin, Gimelshein, Antiga, et~al.]{paszke2019pytorch}
Paszke, A., Gross, S., Massa, F., Lerer, A., Bradbury, J., Chanan, G., Killeen, T., Lin, Z., Gimelshein, N., Antiga, L., et~al.
\newblock {PyTorch}: An imperative style, high-performance deep learning library.
\newblock In \emph{Advances in Neural Information Processing Systems (NeurIPS)}, volume~32, pp.\  8024--8035, 2019.

\bibitem[Shankar et~al.(2022)Shankar, Garcia, Hellerstein, and Parameswaran]{shankar2022operationalizing}
Shankar, S., Garcia, R., Hellerstein, J.~M., and Parameswaran, A.~G.
\newblock Operationalizing machine learning: An interview study.
\newblock \emph{arXiv preprint arXiv:2209.09125}, 2022.

\bibitem[Weng et~al.(2022)Weng, Xiao, Yu, Wang, Wang, He, Li, Zhang, Lin, and Ding]{weng2022mlaas}
Weng, Q., Xiao, W., Yu, Y., Wang, W., Wang, C., He, J., Li, Y., Zhang, L., Lin, W., and Ding, Y.
\newblock {MLaaS} in the wild: Workload analysis and scheduling in large-scale heterogeneous {GPU} clusters.
\newblock In \emph{19th USENIX Symposium on Networked Systems Design and Implementation (NSDI 22)}, pp.\  945--960, 2022.

\bibitem[Yoo et~al.(2003)Yoo, Jette, and Grondona]{yoo2003slurm}
Yoo, A.~B., Jette, M.~A., and Grondona, M.
\newblock {SLURM}: Simple linux utility for resource management.
\newblock In \emph{Job Scheduling Strategies for Parallel Processing (JSSPP)}, Lecture Notes in Computer Science, pp.\  44--60. Springer, 2003.

\bibitem[Zaharia et~al.(2018)Zaharia, Chen, Davidson, et~al.]{zaharia2018mlflow}
Zaharia, M., Chen, A., Davidson, A., et~al.
\newblock Accelerating the machine learning lifecycle with mlflow.
\newblock \emph{IEEE Data Engineering Bulletin}, 41\penalty0 (4):\penalty0 39--45, 2018.

\end{thebibliography}



\end{document}